
%
%
%
%
%
\font\bigrm=cmb10 scaled\magstep1
\font\srm=cmr9
\font\csc=cmcsc10

\def\foot{\baselineskip=12pt\srm\footnote}


\magnification=\magstep1
\hoffset-.27truein
\vsize 8.9truein    \hsize 6.8truein
\topskip 10pt       \leftskip 0pt  \rightskip 0pt
\baselineskip=19pt plus 2pt minus 1pt
\parskip 0pt plus 1pt    \parindent 20pt


\def\->{\rightarrow}     \def\<-{\leftarrow}
\def\<{\langle}          \def\>{\rangle}
\def\[{\left [}          \def\]{\right ]}
\def\({\left (}          \def\){\right )}
\def\|{\vert}

\def\Smat#1{\left[\matrix{#1}\right]}
\def\diff#1{{\partial \over \partial #1}}

\def\hb{\hfill\break}


\def\a{\alpha}
\def\b{\beta}
\def\d{\delta}	    \def\D{\Delta}
\def\e{\epsilon}

\def\la{\lambda}

\def\r{\rho}                          \def\vr{\varrho}
\def\sig{\sigma}    
	        \def\vt{\vartheta}


\def\bZ{{\bf  Z}} \def\bC{{\bf  C}} 
\def\cA{{\cal A}} \def\cF{{\cal F}} \def\cG{{\cal G}} \def\cH{{\cal H}}


\def\da{a^{\dag}}
\def\tr{\tilde\r}\def\tsig{\tilde\sig}\def\tpi{\tilde\pi}

\def\:{\,{ }^\circ_\circ\,}

\def\Hn{\cH_n}\def\Mn{ \cA_q(Mat_n) }
\def\sln{ U_q(sl_n) }\def\slnKM{ U_q(\widehat{sl_n}) }
\def\slKM{ U_q(\widehat{sl_2}) }

\def\xiL#1#2{\xi^{1\,\cdots\,#1}_{1\,\cdots\,#2}\,}

\def\HR{Heisenberg realization}
\def\FFR{free field realization}

\def\Fha{e-mail address : awata@ps1.yukawa.kyoto-u.ac.jp}
\def\Fmn{e-mail address : noumi@tansei.cc.u-tokyo.ac.jp}
\def\Fso{e-mail address : odake@jpnyitp.yukawa.kyoto-u.ac.jp}

\rightline{\vbox{\hbox{YITP/K-1016}
                 \hbox{May 1993}
}}

\vskip.6in\centerline
{\bigrm {\HR} for $\sln$ on the flag manifold }
\vskip.5in\centerline{
{\csc Hidetoshi AWATA}{\foot{$^1$}\Fha} ,
{\csc Masatoshi NOUMI}{\foot{$^2$}\Fmn}  \ and
{\csc Satoru    ODAKE}{\foot{$^3$}\Fso}
}

\it\vskip.3in
\centerline{$^1$ Yukawa Institute for Theoretical Physics}
\centerline{Kyoto University, Kyoto 606, Japan}
\vskip.1in
\centerline{$^2$ Department of Mathematical Sciences, University of Tokyo}
\centerline{Komaba 3-8-1, Meguro-ku, Tokyo 153, Japan}
\vskip.1in
\centerline{$^3$ Department of Physics, Faculty of Liberal Arts}
\centerline{Shinshu University, Matsumoto 390, Japan}

\rm
\vskip.45in\centerline{\bf Abstract}\vskip.25in

 We give the {\HR} for the quantum algebra $\sln$,
which is written by the $q$-difference operator on the flag manifold.
We construct it from the action of $\sln$
on the $q$-symmetric algebra $\Mn$ by the Borel-Weil like approach.
Our realization is  applicable to the construction of
the {\FFR} for the $\slnKM$ [AOS].

\vskip.3in
hep-th/9306010

\vfill\eject
\vskip3mm
\noindent{\bf 1.Introduction }
\vskip2mm

 Recently,
the quantum Knizhnik-Zamolodchikov equations ($q$-KZ eq.) [Sm, FR]
have been analyzed [M1, R].
This $q$-KZ equations are important both for physics and mathematics by
the relationship with 2-dimensional integrable theories [Sm, DFJMN],
quantum affine Lie algebras and elliptic $R$-matrices [FR, DJO].

To solve the classical $(q=1)$ KZ equations,
an important and powerful tools were the {\FFR}
for the affine Lie algebra $\widehat\cG$ [W, FF]
and the {\HR} for the corresponding Lie algebra $\cG$
which is written by the differential operator on the flag manifold
[SV, ATY, FM].
Even in the quantum case $(q \neq 1)$,
for example for the algebra $\slKM$,
the {\HR} and the {\FFR} [FJ, M2, ABG, Sh]
are also important for the analysis of the $q$-KZ equation
[JMMN, KQS, M3].
We expect that this situation is the same for
other quantum affine Lie algebras.

 The aim of this paper is
to construct the {\HR} for the quantum algebra $\sln$.
In the forthcoming paper [AOS],
the {\FFR} for the quantum affine algebra $\slnKM$
will be constructed by using this {\HR}.

\vskip 3mm
\noindent{\bf  2. Quantum algebra $\sln$ }
\vskip2mm

\noindent{\bf \S$\,$2.1.}~
 First we fix some notations.
The algebra $\sln$ is generated by
$e_i$, $f_i$ and invertible $k_i$ $(1\leq i\leq n-1)$ with relations
$$\eqalign{
k_i e_j k_i^{-1} &= q^{ A_{ij}} e_j ,\cr
k_i f_j k_i^{-1} &= q^{-A_{ij}} f_j ,\cr
e_i f_j -f_j e_i &= \d_{ij}{k_i - k_i^{-1} \over q-q^{-1}},\cr
}\qquad\eqalign{
\sum_{m=0}^{1-A_{ij}}(-1)^m \Smat{ 1-A_{ij} \cr m \cr }
e_i^{1-A_{ij}-m} e_j e_i^m &=0 ,\cr
\sum_{m=0}^{1-A_{ij}}(-1)^m \Smat{ 1-A_{ij} \cr m \cr }
f_i^{1-A_{ij}-m} f_j f_i^m &=0 ,\cr
}$$
where $q \in \bC$,
$(A_{ij})_{1\leq i,j\leq n-1}$ is the Cartan matrix such that
$A_{ij}=2\d_{ij}-\d_{i\,j+1}-\d_{i\,j-1}$,
$\big[{n\atop m}\big]=[n]!/[n-m]! [m]!$ and
$[n]=(q^n-q^{-n})/(q-q^{-1})$.


The algebra $\sln$ is a Hopf algebra
with the comultiplication $\D$
$$
\D(k_i)=k_i\otimes k_i                        ,\qquad
\D(e_i)=e_i\otimes 1        + k_i \otimes e_i ,\qquad
\D(f_i)=f_i\otimes k_i^{-1} + 1   \otimes f_i ,
$$
the antipode $S$ such that
$S(k_i)= k_i^{-1}$, $S(e_i)=-k_i^{-1}e_i$, $S(f_i)=-f_i k_i$
and the co-unit $\e$ such that
 $\e(k_i)=1$, $\e(e_i)=0$, $\e(f_i)=0$.

\vskip 2mm
\noindent{\bf \S$\,$2.2.}~
 Let $M_{\la}$ be the Verma module over $\sln$
generated by the highest weight vector $\|\la\>$ such that
$e_i\|\la\> = 0 $,
$k_i\|\la\> = q^{\la_i}\|\la\>$ with $\la_i\in\bC$.
 The dual module $M_{\la}^*$ is generated by $ \<\la\|$ which satisfies
$\<\la\|f_i = 0 $,
$\<\la\|k_i = q^{\la_i}\<\la\|$.
The bilinear form
$M_{\la}^* \otimes M_{\la} \-> \bC  $ is uniquely defined by
$\<\la\|\la\> = 1 $ and
$\big(\<u\|X\big)\|v\> = \<u\|\big(X\|v\>\big) $
for any $\<u\|\in M_{\la}^*$,  $\|v\>\in M_{\la}$ and $X\in\sln$.

\vskip 3mm
\noindent{\bf  3. {\HR} for $\sln$ }
\vskip2mm

\noindent{\bf \S$\,$3.1.}~
The Heisenberg algebra $\Hn$ is generated by
the coordinate $x_{ij}$, $x_{ij}^{-1}\in\bC$ and
the differential operator $\vt_{ij}=x_{ij}\diff{x_{ij}}$
($1\leq i<j\leq n$) with relation
$[\vt_{ij},x_{kl}] = \d^i_k \d^j_l x_{kl}$
or equivalently
$$
q^{ \vt_{ij} } x_{kl} q^{ -\vt_{ij} } =  q^{\d^i_k \d^j_l} x_{kl}.
$$

The quantum algebra $\sln$ is realized by the Heisenberg algebra $\Hn$.
We have \hb
\noindent{\underbar{\bf Theorem I}}.~{\it
There exists the algebra homomorphism $\pi_{\la}:\sln\->\Hn$ define as
$$\eqalign{
\pi_\la (k_i)&=
q^{\sum_{j=1  }^{i-1}( \vt_{ji  }- \vt_{j\,i+1} )
                    +( \la_i     -2\vt_{i\,i+1} )
  +\sum_{j=i+2}^n    ( \vt_{i+1\,j}- \vt_{ij  } )  } ,\cr
\pi_\la (e_i)&=\sum_{k=1}^i
q^{\sum_{j=1  }^{k-1}( \vt_{ji  }-\vt_{j\,i+1} )  }
  {x_{ki  } \over x_{k\,i+1} }[\vt_{k\,i+1}] ,\cr
\pi_\la (f_i)&=\sum_{k=1}^{i-1}
  {x_{k\,i+1} \over x_{ki  } }[\vt_{ki  }]
q^{-\sum_{j=k+1}^{i-1}( \vt_{ji  }- \vt_{j\,i+1} )
                     -( \la_i     -2\vt_{i\,i+1} )
   -\sum_{j=i+2}^n    ( \vt_{i+1\,j}- \vt_{ij  } )  }  \cr
&+ x_{i\,i+1} [         ( \la_i     - \vt_{i\,i+1} )
   +\sum_{j=i+2}^n    ( \vt_{i+1\,j}- \vt_{ij  } )  ]  \cr
&- \sum_{k=i+2}^{n}{x_{ik} \over x_{i+1\,k} }[\vt_{i+1\,k  }]
q^{                     \la_i
   +\sum_{j=k  }^n    ( \vt_{i+1\,j}- \vt_{ij  } )  },  \cr
}$$
with $x_{ii}=1$.\hb
}
Here $[n]$ denotes the $q$ integer,
so $\pi_{\la} (g)$'s are the $q$-difference operators.
The proof will be given in the next section.

\def\Fdual{
These dual generators relate to
the screening currents of the {\FFR} for $\slnKM$ [AOS]
which must important to the analysis of the $q$-KZ equation.
}

We also have the following dual generators{\foot\dag\Fdual}

\noindent{\underbar{\bf Theorem II}}.~{\it
There exists the algebra anti-homomorhpism $\tpi_{\la}:\sln\->\Hn$,
$\tpi_{\la}=\tsig\circ\pi_{\la}\circ\sig$, with $\sig$ such that
$\sig(k_i) =k_{n-i}$,
$\sig(e_i) =e_{n-i}$,
$\sig(f_i) =f_{n-i}$ and $\tsig$ such that
$\tsig(  x_{ij}) =    x_{n+1-j\;n+1-i}$,
$\tsig(\vt_{ij}) = -\vt_{n+1-j\;n+1-i}$,
$\tsig(\la_{i }) = -\la_{      n+1-i}$.
}

\vskip 2mm
\noindent{\bf \S$\,$3.2.}~
 Let $\cF=\bC [x_{ij}]\,\|0\>$
be the Fock module over Heisenberg algebra $\Hn$
generated by the highest weight vector $\|0\>$ such that
$x_{ij}^{-1}\|0\> = \vt_{ij}\|0\> = 0 $.
 The dual module $\cF^*=\<0\|\,\bC[x_{ij}^{-1}]$
is generated by $\<0\|$ which satisfies
$\<0\|x_{ij} = \<0\|\vt_{ij} = 0 $.
The bilinear form
$\cF^* \otimes \cF \-> \bC  $ is uniquely defined by
$\<0\|0\> = 1 $ and
$\big(\<u\|X\big)\|v\> = \<u\|\big(X\|v\>\big) $
for any $\<u\|\in \cF^*$,  $\|v\>\in \cF$ and $X\in\cH_n$.
For $\<0\| f(x_{ij}^{-1}) \in \cF^* $ and $ g(x_{ij})\|0\>\in \cF $,
$\<0\| f(x_{ij}^{-1}) \, g(x_{ij})|0\>$
is nothing but the constant part of $f(x_{ij}^{-1}) \, g(x_{ij})$.

\vskip 3mm
\noindent{\bf  4. Construction of the {\HR} for $\sln$}
\vskip2mm

Next we prove above Theorems by a Borel-Weil like approach,
which is based on the method in Ref. [N].
First we give some notations.

\vskip 2mm
\noindent{\bf \S$\,$4.1.}~
The $q$-symmetric algebra $\Mn$ is generated by
$t_{ij}$ $(1\leq i,j\leq n)$ with relations
$$
\eqalign{
t_{ik}t_{jk}&=qt_{jk}t_{ik}, \cr
t_{ik}t_{il}&=qt_{il}t_{ik}, \cr
}\qquad\eqalign{
t_{il}t_{jk}&= t_{jk}t_{il}, \cr
t_{ik}t_{jl} -qt_{il}t_{jk}&=t_{jl}t_{ik} -q^{-1}t_{jk}t_{il},
}\leqno{(4.1)}$$
for $i<j$ and $k<l$.
Note that this algebra has the algebra automorphism $\r$ such that
$\r(t_{ij})=t_{n+1-j\;n+1-i}$, $\r(q)=q^{-1}$
and the algebra anti-automorphism $\tr$ such that
$\tr(t_{ij})=t_{n+1-j\;n+1-i}$, $\tr(q)=q$.

The algebra $\Mn$ has the structure of a $\sln$-module.
The action of $\sln$ on $\Mn$ is
$$\eqalign{
k_m t_{ij}= t_{ij  } q^{\d_{mj} - \d_{m+1\,j}}&,\qquad
e_m t_{ij}= t_{i\,j-1}            \d_{m+1\,j} ,\qquad
f_m t_{ij}= t_{i\,j+1}  \d_{mj},\cr
g(uv) &=\sum_a (g^{'}_a u) (g^{''}_a v),\qquad
g.1=\e(g)1,
}$$
for all $u$, $v\in\Mn$ and
for all $g\in\sln$ with $\D(g)=\sum_a g^{'}_a \otimes g^{''}_a$.
Note that this action of $g\in\sln$ can be written by
the matrix $\vr(g)_{ij}$ as
$g\,t_{ij} = \sum_k t_{jk}\vr(g)_{kj}$ with
$\vr(k_m) =q^{E_{mm}-E_{m+1\,m+1}}$,
$\vr(e_m) =E_{m\,m+1}$,
$\vr(f_m) =E_{m+1\,m}$ and
$(E_{\a\b})_{ij}=\d_{\a i}\d_{\b j}$.
These matrices are noting but the vector representation for the $\sln$.
The action for the rows of matrix $t_{ij}$
is given by the above automorphism $\r$ or $\tr$.

\vskip 2mm
\noindent{\bf \S$\,$4.2.}~
For the ordered set
$I=\{ i_1 < \cdots < i_r \}$ and $J=\{ j_1 < \cdots < j_r \}$,
let  $\xi^I_J$ be the quantum $r$-minor determinant
with respect to rows $I$ and columns $J$ such that [TT, NYM]
$$
\xi^I_J=\sum_{\sig\in{\bf S}_r}(-q)^{l(\sig)}
t_{i_{\sig(1)}j_1}\cdots t_{i_{\sig(r)}j_r}.
$$
Here ${\bf S}_r$ is the permutation group of the set $\{1,\cdots,r\}$
and $l(\sig)$ stands for the number of inversions involved in $\sig$ ;
$l(\sig)=\sharp \{ (i,j)\, ; \, i<j,\, \sig(i) > \sig(j) \}$.
{}From now on, $\xi^I_J=0$ if $I$ or $J$ has same elements.
Note that $\xi^I_J\xi^{I'}_{J'}=\xi^{I'}_{J'}\xi^I_J$
if $I'\subset I$, $J'\subset J$. We have

\noindent{\underbar{\bf Proposition}}.~{\it
With the lower triangular matrix $B$,
the Gauss decomposition of
the matrix $T=(t_{ij})$ of the $q$-coordinates is given as
$$
t_{ij}=\sum_k B_{ik}X_{kj},\qquad
B_{ij}= (\xiL{j-1}{j-1})^{-1} \xiL{j-1\,i}{j     } ,\qquad
X_{ij}= (\xiL{i  }{i  })^{-1} \xiL{i     }{i-1\,j} ,
$$
and $B_{ij}=0$ for $i<j$ and $X_{ij}=0$ for $i>j$.
Here $\{1\cdots 0\} =\{ \; \}$.
}

\noindent{\it Proof}.~
follows from
$$
t_{ij}=B_{i1}X_{1j}+(\xi^1_1)^{-1}\xi^{1i}_{1j},\qquad
(\xiL{r  }{r  })^{-1}\xiL{r  \,i}{r  \,j}=B_{i\,r+1}X_{r+1\,j}+
(\xiL{r+1}{r+1})^{-1}\xiL{r+1\,i}{r+1\,j},
$$
which are obtained from the $q$-deformed Jacobi identity
$$
   \xiL{r     }{r     } \xiL{r\,r+1\,r+2}{r\,r+1\,r+2}
=  \xiL{r\,r+1}{r\,r+1} \xiL{r\,     r+2}{r\,     r+2}
-q \xiL{r\,r+1}{r\,r+2} \xiL{r\,     r+2}{r\,r+1     }.
\eqno{\rm Q.E.D.}$$

We regard $X_{ij}$ $(i<j)$ as a $q$-analogue of local coordinates of
the flag manifold $B\backslash GL_n$.
For $i<i_1$ and $ I=\{i_1<\cdots<i_r\}$,
we denote $\eta^i_{I}=\xiL{i\,i+1\cdots i+r}{i\,i_1\cdots i_r}$,
then $X_{ij}=(\eta^{i-1}_i)^{-1}\eta^{i-1}_j$.
Since the principal minors $\xiL{i}{i}$'s $1\leq i\leq n$
commute with each other,
one can consistently adjoin their inverse to the algebra $\bC[\xi^I_J]$.

\vskip 2mm
\noindent{\bf \S$\,$4.3.}~
The quantum minor $\eta^r_{ij}$'s satisfy, for $r<i<j<k<l$,
the same relations as $t_{ij}$'s in (4.1)
and Pl\"uker relation ( Young symmetry ) [TT, NYM, N]
$$
\eta^r_i\eta^r_{jk}-q\eta^r_j\eta^r_{ik}+q^2\eta^r_k\eta^r_{ij}=0,
$$
and the commutation relations
$$\eqalign{
\eta^r_{i }\eta^r_{jk}=q  \eta^r_{jk}&\eta^r_{i },\qquad
\eta^r_{ij}\eta^r_{k }=q  \eta^r_{k }\eta^r_{ij},\qquad
\eta^r_{ik}\eta^r_{j }=   \eta^r_{k }\eta^r_{ij}+\eta^r_{i }\eta^r_{jk},\cr
&\eta^r_{ij}\eta^r_{jk}=q \eta^r_{jk}\eta^r_{ij},\qquad
\eta^r_{ij}\eta^r_{kl}=q^2\eta^r_{kl}\eta^r_{ij}.
}$$

The action of $\sln$ on the quantum minor $\eta^i_j$ is
$$\eqalign{&
k_m \eta^{i-1}_{i  j  }
  = \eta^{i-1}_{i  j  }q^{\d_{mj} -\d_{m+1\,j} +\d_{mi} },\cr
e_m \eta^{i-1}_{i  j  }
  = \eta^{i-1}_{i\,j-1}&\d_{m+1\,j} ,\qquad
f_m \eta^{i-1}_{i  j  }
  = \eta^{i-1}_{i\,j+1}\d_{mj}
  + \eta^{i-1}_{i+1\,j}\d_{mi}.
}$$
Owning to the Pl\"uker relation,
$ \eta^{i-1}_{i+1\,j}=
  \eta^{i-1}_{i+1}(\eta^{i-1}_i)^{-1}\eta^{i-1}_{ij}
-q\eta^{i-1}_{j  }(\eta^{i-1}_i)^{-1}\eta^{i-1}_{i\,i+1}$,
the algebra $\cA=\bC[\eta^{i-1}_j,(\eta^{i-1}_i)^{-1}]
_{1\leq i\leq n-1,\;i\leq j\leq n}$
has the structure of a $\sln$-module.

\vskip 2mm
\noindent{\bf \S$\,$4.4.}~
To relate the non-commutative algebra $\bC[X_{ij}]$
with the commutative one $\bC[x_{ij}]$,
we fix the ordering of $\eta^i_j$'s.
The algebra $\cA$ has the basis
$$\{\;
(\eta^{0}_n)^{a_{1n}}\cdots(\eta^{0  }_1    )^{a_{11}}
(\eta^{1}_n)^{a_{2n}}\cdots(\eta^{1  }_2    )^{a_{22}}
                     \cdots(\eta^{n-2}_{n-1})^{a_{n-1\,n-1}}\;|\;
a_{ij}\in\bZ_{\geq 0},\;i<j,\;  a_{ii}\in\bZ\;\},
$$
which ordering we call {\it normal ordering}.
We introduce the projection $\:*\: :\cA\->\cA$ such that
$$
\:  any\; ordered\; \prod_{i\leq j}(\eta^{i-1}_{j})^{a_{ij}}\:
= normal\; ordered\; \prod_{i\leq j}(\eta^{i-1}_{j})^{a_{ij}}.
$$

Let
$Z_{\la}^a=\:\prod_i(\eta^{i-1}_i)^{\la_i}\prod_{j<k}(X_{jk})^{a_{jk}}\:$
with $\la_i\in\bZ$ and $a_{ij}\in\bZ_{\geq 0}$.
If we denote $Y^i=(\eta^{i-1}_n)^{a_{in}}\cdots(\eta^{i-1}_i)^{a_{ii}}$
with $a_{ii}=\la_i-\sum_{j=i+1}^n a_{ij}$,
then $Z_{\la}^a=Y^1\cdots Y^{n-1}$.
The algebra $\cA$ has the decomposition $\cA=\oplus_{\la_i\in\bZ}\cA_{\la}$
such that $\cA_{\la}$ is the vector space spanned by the vectors
$\{Z_{\la}^a\;|\;a_{ij}\in\bZ_{\geq 0},\;i<j\;\}$.
The algebra $\cA_{\la}$ also has the structure of a $\sln$-module,
and we have

\noindent{\underbar{\bf Lemma}}.~{\it
The left action of $\sln$ on $\cA_{\la}$ is as follows
$$\eqalign{
k_i Z_{\la}^a&= Z_{\la}^a
q^{\sum_{j=1  }^{i-1}(a_{ji}-a_{j\,i+1})+(\la_i-2a_{i\,i+1})
  +\sum_{j=i+2}^n    (a_{i+1\,j}- a_{ij }) },\cr
e_i Z_{\la}^a&=\sum_{k=1}^i
q^{\sum_{j=1  }^{k-1}(a_{ji}-a_{j\,i+1})}[a_{k\,i+1}]
\: Z_{\la}^a (X_{k\,i+1})^{-1} X_{ki} \:,\cr
f_i Z_{\la}^a&=\sum_{k=1}^{i-1}[a_{ki}]
\:X_{k\,i+1} (X_{ki})^{-1} Z_{\la}^a\:
q^{-\sum_{j=k+1}^{i-1}(a_{ji}-a_{j\,i+1})-(\la_i-2a_{i\,i+1})
   -\sum_{j=i+2}^n    (a_{i+1\,j}- a_{ij})}\cr
&+[(\la_i-a_{i\,i+1})+\sum_{j=i+2}^n (a_{i+1\,j}-a_{ij})]
\:X_{i\,i+1} Z_{\la}^a\:\cr
&-\sum_{k=i+2}^{n}[a_{i+1\,k}]\:X_{ik} (X_{i+1\,k})^{-1} Z_{\la}^a\:
q^{\la_i+\sum_{j=k}^n(a_{i+1\,j}-a_{ij})}.\cr
}$$
}

\noindent{\it Proof}.~
follows from
$$\eqalign{
k_m Y^i&=Y^i q^{(a_{im}-a_{i\,m+1})\sum_{j=i}^{n-1}\d_{mj}
  +\d_{m\,i-1}\sum_{j=i+1}^n a_{m+1\,j}},\cr
e_m Y^i&=[a_{i\,m+1}]\:Y^i(\eta^{i-1}_{m+1})^{-1}\eta^{i-1}_{m}\:
\sum_{j=i+1}^{n}\d_{m+1\,j},\cr
f_m Y^i
&=[a_{im}]\:\eta^{i-1}_{m+1}(\eta^{i-1}_m)^{-1}Y^i\:\sum_{j=i}^{n-1}\d_{mj}
+\d_{m\,i-1}\sum_{k=i+1}^n[a_{ik}]\eta^{i-2}_{ik}\:(\eta^{i-1}_k)^{-1}Y^i\:
q^{-\sum_{j=i+1}^{k-1}a_{ij}}\cr
&=[a_{im}]\:\eta^{i-1}_{m+1}(\eta^{i-1}_m)^{-1}Y^i\:\sum_{j=i}^{n-1}\d_{mj}
+\d_{m\,i-1}[\sum_{k=i+1}^n a_{ik}]
\:\eta^{i-2}_i(\eta^{i-2}_{i-1})^{-1}Y^i\:\cr &
-\d_{m\,i-1}\sum_{k=i+1}^n[a_{ik}]
\:\eta^{i-2}_k(\eta^{i-2}_{i-1})^{-1}\eta^{i-1}_k(\eta^{i-1}_i)^{-1}Y^i\:
q^{-\sum_{j=k}^n a_{ij}},
}$$
here we use
$k_m(\eta^r_i)^a =                    (k_m\eta^r_i)^a$,
$e_m(\eta^r_i)^a = [a](\eta^r_i)^{a-1}(e_m\eta^r_i)$,
$f_m(\eta^r_i)^a = [a]                (f_m\eta^r_i) (\eta^r_i)^{a-1}$
with $a\in\bZ$ and the identity
$\sum_k [a_k]q^{\big(\sum_{j<k}-\sum_{j>k}\big)a_j}=[\sum_k a_k]$.
The polynomials of $q$ in $e_i Z_{\la}^a$ and $f_i Z_{\la}^a$
come from the Cartan parts of the comultiplication of $e_i$ and $f_i$
respectively.

\hfill Q.E.D.

\vskip 2mm
\noindent{\bf \S$\,$4.5.}~
{\it Proof of Theorem I}.~\hb\noindent
We consider the commutative algebra $\bC[ x_{ij}]_{1\leq i<j\leq n}$
and define an isomorphism
$\pi_{\la}:\cA_{\la}\->\bC[ x_{ij}]$ by
$\pi_{\la}(Z_{\la}^a)=z^a$, with
$z^a=\prod_{r<j}(x_{rj})^{a_{rj}}$.
Applying this isomorphism $\pi_\la$ to above Lemma,
we obtain the $q$-difference operators on $\bC[ x_{ij}]$ in Theorem I.
\hfill {Q.E.D.}


\vskip 2mm
\noindent{\bf \S$\,$4.6.}~
{\it Proof of Theorem II}.~\hb\noindent
With the lower triangular matrix $\tilde B$,
the Gauss decomposition of inverse direction $T=\tilde X \tilde B$ is
obtained by the algebra anti-automorphism $\tr$ in \S 4.1
from the Gauss decomposition $T=BX$.
By the algebra automorphism $\r$ with some sign changing,
we get the action of $\sln$ on $\bC[\tilde X_{ij}]$
and the dual generators of Theorem II.
\hfill {Q.E.D.}


\bigbreak\noindent{\bf Conclusion and Discussion.}

 We constructed the {\HR} for the $\sln$ by the flag coordinate,
which is applicable to the construction of the {\FFR} for the $\slnKM$ [AOS].
In the Ref. [DJMM], they also gave the similar realization for the $\sln$
but it seems that it can not be affinized.


\bigbreak\noindent{\bf Acknowledgments.}

 The authors would like to thank
E. Frenkel, K. Hasegawa, M. Jimbo, K. Kimura, G. Kuroki,
F. Malikov, A. Matsuo, T. Miwa, J. Shiraishi, Y. Yamada
and the members of KEK, RIMS and YITP for valuable discussions.
H.A and S.O are supported by Soryushi-syougakkai.

\vskip3mm
\noindent{\bf  Appendix. The Jordan-Schwinger type realization
                         and $q$-oscillator}
\vskip2mm

\noindent{\bf \S$\,$A.1.}~
If we consider only $i=1$ of $t_{ij}\in\Mn$,
then we can obtain the $n$ variables
Jordan-Schwinger type realization for $\sln$ [H, Z].
Let us denote $t_j=t_{1j}$,
the algebra $\cA=\bC [t_i]_{1\leq i\leq n}$ has the basis
$\{\; t_n^{a_n}\cdots t_1^{a_1}\;|\;a_i\in\bZ_{\geq 0}\;\}$,
which ordering we call {\it normal ordering},
and has the structure of a $\sln$-module.
By an isomorphism $\pi:\cA\->\bC[x_i]$,
$\pi(t_n^{a_n} \cdots t_1^{a_1})=x_1^{a_1}\cdots x_n^{a_n}$
and by the action of $\sln$ on $\cA$, we obtain

\noindent{\underbar{\bf Proposition}}.~{\it
There exists the algebra homomorphism $\pi:\sln\->\Hn$ define as
$$
\pi(k_i)= q^{\vt_{i}-\vt_{i+1}},\quad
\pi(e_i)={x_{i  }\over x_{i+1}}[\vt_{i+1}],\quad
\pi(f_i)={x_{i+1}\over x_{i  }}[\vt_{i  }].
$$
}

We introduce the projection $\:*\:$ same as before.
Denote $X_i=t_1^{-1}t_i$ $(2\le i\le n)$ and
$Z_{\la}^a=\:t_1^{\la}\prod_{i=2}^n X_i^{a_i}\:=t_n^{a_n}\cdots t_1^{a_1}$
with $a_1=\la-\sum_{i=2}^n a_i$,
then the algebra $\cA[t_1^{-1}]$ has the decomposition
$\cA[t_1^{-1}]=\oplus_{\la\in\bZ}\cA_{\la}$
such that $\cA_{\la}$ is the vector space spanned by the vectors
$\{Z_{\la}^a\;|\;a_{i}\in\bZ_{\geq 0},\;i>1\;\}$.
By an isomorphism $\pi_{\la}:\cA_{\la}\->\bC[x_i]$,
$\pi_{\la}(t_n^{a_n} \cdots t_1^{a_1})=x_2^{a_2}\cdots x_n^{a_n}$
and by the action of $\sln$ on $\cA_{\la}$,
we obtain the $n-1$ variables inhomogeneous realization for $\sln$,
which is the same as above Proposition with additional conditions
$x_1=1$ and $\vt_1=\la-\sum_{i=2}^n\vt_i$.
This realization corresponds with that in Theorem-I
on $\bC[x_{1j}]$ with $\la_i=0$ for $i\neq 1$.

\vskip 2mm
\noindent{\bf \S$\,$A.2.}~
For the Heisenberg algebra $\< x,\vt\>$ with $q^{\vt}xq^{-\vt}=qx$,
if we denote
$$
a=x,\quad \da={1\over x}[\vt],\quad  N=\vt,
$$
then $\< a,\da,N\>$ satisfies the $q$-oscillator algebra such that
$$
a \da=[N],\quad \da a=[N+1],
$$
which is equivalent to $\da a-q^{\pm 1}a\da=q^{\mp N}$.
And they satisfy $[N,a]=a$, $[N,\da]=-\da$.

So we can rewrite our Theorem by the $q$-oscillator algebra.

\vfill\eject
\noindent{\bf References}
\vskip2mm
\item{ }{ }
\vskip-0.5truecm
\baselineskip=14pt

\itemitem{[ABG]}{A. Abada, A. Bougourzi and M. El Gradechi,
{\it Deformation of the Wakimoto construction,}
   (hep-th/9209009) preprint CRM-1829 (1992).}

\itemitem{[AOS]}{H. Awata, S. Odake and J. Shiraishi,
{\it Free Boson representation of $U_q(\widehat{sl_3})$,}
   (hep-th/9305017) preprint RIMS-920,YITP/K-1017;
{\it Free Boson realization of $U_q(\widehat{sl_N})$,}
   (hep-th/9305146) preprint RIMS-924,YITP/K-1018}

\itemitem{[ATY]}{H. Awata, A. Tsuchiya and Y. Yamada,
{\it Integral formulas for the WZNW correlation functions,}
   Nucl. Phys. {\bf B365} (1991) 680-696.\hfill\break
   H. Awata,
{\it Screening current Ward identity and
   integral formulas for the WZNW correlation functions,}
   Prog. Theor. Phys. Supplement {\bf 110} (1992) 303-319.}



\itemitem{[DFJMN]}{B. Davies, O. Foda, M. Jimbo, T. Miwa and A. Nakayashiki,
{\it Diagonalization of the XXZ Hamiltonian by vertex operators,}
   Commun. Math. Phys. {\bf 151} (1993) 89-154.}

\itemitem{[DJMM]}{E. Date, M. Jimbo, K. Miki and T. Miwa,
{\it Cyclic representations of $U_q(sl(n+1,\bC))$ at $q^N =1$,}
   Publ. RIMS. Kyoto Univ. {\bf 27} (1991) 347-366.}

\itemitem{[DJO]}{E. Date, M. Jimbo and M. Okado,
{\it Crystal base and $q$-vertex operators,}
   Osaka Univ. Math. Sci. preprint {\bf 1} (1991).}

\itemitem{[FF]}{B. Feigin and E. Frenkel,
{\it A family of representations of affine Lie algebras,}
   Russian Math. Surveys {\bf 43} (1988) 221-222;
{\it Affine Kac-Moody algebras and semi-infinite Flag manifolds,}
   Commun. Math. Phys. {\bf 128} (1990) 161-189.}

\itemitem{[FJ]}{I. Frenkel and N. Jing,
{\it Vertex representations of quantum affine algebras,}
   Proc. Nat'l. Acad. Sci. USA {\bf 85} (1988) 9373-9377.}

\itemitem{[FM]}{B. Feigin and F. Malikov,
{\it Integral intertwining operators and
   complex powers of differential ( $q$-difference ) operators,}
   preprint RIMS-894 (1992).}

\itemitem{[FR]}{I. Frenkel and N. Reshetikhin,
{\it Quantum affine algebras and holonomic difference equations,}
   Commun. Math. Phys. {\bf 146} (1992) 1-60.}

\itemitem{[H]}{T. Hayashi,
{\it $Q$-analogues of Clifford and Weyl algebras -
   Spinor and Oscillator representations of quantum enveloping algebras,}
   Commun. Math. Phys. {\bf 127} (1990) 129-144.}

\itemitem{[JMMN]}{M. Jimbo, K. Miki, T. Miwa and A. Nakayashiki,
{\it Correlation functions of the XXZ model for $\D < -1$,}
   Phys. Lett. {\bf A168} (1992) 256-263.}

\itemitem{[KQS]}{A. Kato, Y. Quano and J. Shiraishi,
{\it Free Boson Representation of $q$-Vertex Operators
     and their Correlation Functions,}
     Tokyo Univ. preprint  UT-618 (1992).}

\itemitem{[M1]}{A. Matsuo,
{\it Jackson integrals of Jordan-Pochhammer type
   and quantum Knizhnik-Zamolodchikov equations,}
   Nagoya Univ. preprint (1992);
{\it Quantum algebra structure of certain Jackson integrals,}
   Nagoya Univ. preprint (1992).  }

\itemitem{[M2]}{A. Matsuo,
{\it Free field representation of
    quantum affine algebra $U_q(\widehat{sl_2})$,}
   (hep-th/9208079) Nagoya Univ. preprint (1992).}

\itemitem{[M3]}{A. Matsuo,
{\it Free field representation of
   $q$-deformed primary fields for $U_q(\widehat{sl_2})$,}
   (hep-th/9212040) Nagoya Univ. preprint (1992).}

\itemitem{[N]}{M. Noumi,
{\it Quantum Grassmannians and $q$-hypergeometric series,}
   CWI Quarterly {\bf 5} (1992) 293-307.}

\itemitem{[NYM]}{M. Noumi, H. Yamada and K. Mimachi,
{\it Finite dimensional representations of the quantum group $GL_q(n;\bC)$
   and the zonal spherical functions on $U_q(n-1)\backslash U_q(n)$,}
   to appear in Japanese J. Math.}

\itemitem{[R]}{N. Reshetikhin,
{\it Jackson-type integrals, Bethe vectors, and solutions
   to a difference analog of the Knizhnik-Zamolodchikov system,}
   Lett. Math. Phys. {\bf 26} (1992) 153-165.}

\itemitem{[Sh]}{J. Shiraishi,
{\it Free boson representation of $U_q(\widehat{sl_2})$,}
   Phys. Lett. {\bf A171} (1992) 243-248.}

\itemitem{[Sm]}{F. Smirnov,
{\it Dynamical symmetries of massive integrable models,}
   Int. J. Mod. Phys. {\bf A7} Supplement 1 (1992) 813-837;839-858.}

\itemitem{[SV]}{V. Schechtman and A. Varchenko,
{\it Arrangements of hyperplanes and Lie algebra homology,}
   Invent. Math. {\bf 106} (1991) 139-194.}

\itemitem{[TT]}{E. Taft and J. Towber,
{\it Quantum deformation of flag schemes and Grassmann schemes I,}
   J, Algebra {\bf 142} (1991) 1-36.}

\itemitem{[W]}{M. Wakimoto,
{\it Fock representations of the affine Lie algebra $A_1^{(1)}$,}
   Commun. Math. Phys. {\bf 104} (1986) 605-609.}

\itemitem{[Z]}{see references in C. Zachos,
{\it Paradigms of quantum algebras,}
   Symmetries in Science {\bf V},
   B. Gruber et al. (eds.), Plenum, 1991, p.593-609.}

\bye